\documentstyle[preprint,aps]{revtex}
\tighten
\begin{document}
\draft
\preprint{MA/UC3M/4/93}
\title{Phase transitions in two-dimensional traffic flow models}
\author{Jos\'e A.\ Cuesta, Froil\'an C.\ Mart\'\i nez, Juan M.\ Molera,
and Angel S\'anchez}
\address{Escuela Polit\'{e}cnica Superior, Universidad Carlos III de
Madrid, \\
Avda.\ Mediterr\'{a}neo 20, E-28913 Legan\'{e}s, Madrid, Spain}
\date{\today}
\maketitle
\begin{abstract}
We introduce two simple two-dimensional lattice models to study
traffic flow in cities. We have found that a few basic elements give
rise to the characteristic phase diagram of a first-order phase
transition from a freely moving phase to a jammed state, with a
critical point. The jammed phase presents new transitions
corresponding to structural transformations of the jam. We discuss
their relevance in the infinite size limit.
\end{abstract}

\vspace{1cm}


\narrowtext

Car displacements through large cities, data exchange among processors
in a massively parallel computer, or communications in computer
networks, are examples of situations where avoiding undesirable traffic
jams is extremely important \cite{Libros}. It is therefore necessary to
achieve a comprehensive understanding of the mechanisms leading to the
nucleation, growth, and evolution of these traffic jams, as they can be
responsible for decrease or even full suppression of flow between
different parts of the system. To this end, models based on cellular
automata (CA) seem to be a suitable tool to approach the problem
because of both their nature and their computational efficiency.
Some one-dimensional \cite{Libros,2,3,4} and two-dimensional \cite{5}
models have been proposed in the past to study different traffic flow
problems. In the one-dimensional case, there is good agreement between
CA \cite{4} and fluid-dynamical \cite{6} results, as well as to data
obtained from actual traffic in highways \cite{Libros,4}. Thus, the
main aspects of car flow in freeways or individual streets seem to be at
least qualitatively understood. As regards 2D CA models, the only ones
we are aware of are based upon the fundamental assumption that cars
never turn \cite{5}.
The traffic-jam transition reported in \cite{5} can be a direct
consequence of this unrealistic hypothesis.
In this Letter we show that allowing cars to turn (as in actual
situations) brings along
the appearance of a complex phase diagram.

Our model consists of the following basic ingredients.
We have cars moving inside a town. The town is made of
one-way perpendicular ($L$ horizontal and $L$ vertical) streets
arranged in a square lattice with periodic boundary conditions. The way
of every street is fixed independently according to a certain rule that
depends on the particular choice of the model (see below). Cars
sit at the crossings, and they can move to one of its nearest neighbors
(allowed by the direction of the streets) every time step. Two cars
cannot be at the same crossing simultaneously. Each car is
assigned a {\em trend} or preferred direction, ruled
by a variable $w_i({\bf r})$, which we define as the probability that
car $i$, located at node ${\bf r}$, jumps to the allowed neighbor in
the horizontal street (accordingly, $1-w_i({\bf r})$ is the probability
to move vertically).
Finally, there are traffic lights, that
permit horizontal motion at even time steps and vertical motion at odd
time steps.

We now define the dynamics of the model. Every time step
$w_i({\bf r})$ is evaluated and the direction
where to move next is chosen accordingly. Then, it is checked
that the chosen site is empty
and that the motion is allowed by the light; otherwise the car will not
be moved. Finally,
all cars that can be moved are placed at their destination
site and the next
time step starts. We want to stress that the whole process is carried
out simultaneously for all cars. The fact that traffic lights allow
motion alternatively in vertical and horizontal streets prevents two
cars from colliding at any crossing.

In this Letter, we concern ourselves with two of the simplest versions
of the model, that we name model A and model B.
Both models are characterized by a unique parameter $\gamma$
which we call {\em randomness}. This parameter allows us to control the
trend of the motion of every car by defining $w_i({\bf r})$ through it.
Model A has streets pointing only up and left. Half of the cars are
given a trend $w_i({\bf r})=\gamma$, and the other half are set to
$w_i({\bf r})=1-\gamma$. This amounts to having half of the cars moving
preferently upwards and the other half leftwards. For simmetry reasons,
it is enough to study the range $0\leq\gamma\leq 1/2$.
It has to be noticed that, in case
we fix $\gamma=0$, the cars are deterministic and
always move along their preferred
direction, and Model A becomes Model I of Ref.\ \cite{5}.
Model B is defined in the following way. It has
streets that point alternatively up and down, and right and left.
We work with four equal-number groups of cars: Each of them is
assigned one of the four possible directions as its trend. This is
accomplished by the following definition: Upward or downward bound cars
have
$w_i({\bf r})=\gamma$ if a street with the same direction as the
trend of the $i$-th car
passes through site {\bf r}, and $w_i({\bf r})=1-\gamma$ otherwise; left
or right cars behave the other way around.

On the above described models, we have carried out an extensive
simulation program, simulating the corresponding CA's on towns of
32$\times$32, 64$\times$64, and 128$\times$128 sites. A typical run
consists of the evolution along 10$^6$ time steps of a randomly chosen
initial condition for a given density (number of cars/number of lattice
sites). For every time step we monitor the {\em mean} velocity,
defined as the number of moved cars divided by the total number
of cars. By means of this magnitude, we distinguish when the system
reaches a steady state. Once in this state, we perform time averages on
magnitudes of interest until the end of the simulation.
We have also studied the outcome of different randomly chosen initial
configurations. Although in general, these outcomes are similar,
a very particular dependence is found in some parts of
the phase diagram (see below). We have also checked different random
number
generators \cite{NumRec} and all of them
lead to the same results. We have studied these models
for a number of car densities ranging from $n=0$ to $n=1$, and for
randomness values on the range $0\leq\gamma\leq 1/2$.
In addition, we have recorded any possible structure of the traffic
jam in the steady state by measuring the average
occupation time per site, defined as the number of time steps during
which a site is occupied by a {\em stopped} car divided by the total
averaging time. All simulations were performed in workstations HP 720,
and DEC 3100 and 5100; a typical simulation for a given car density on
a 64$\times$64 town takes about 2 hours of CPU time, and 12 hours for a
128$\times$128 town (this is for model A, for model B times are
approximately a 25\% higher).

Results for model A are summarized in Fig.\ \ref{Isotermas}. Such a
figure can be understood as the phase diagram of a first order phase
transition \cite{stanley} from a freely moving to a jammed phase. The
curves $v(n)$
undergo a discontinuous transition  of magnitude $\Delta v(\gamma)$ at
density $n_t(\gamma)$. As $\gamma$ increases, $n_t(\gamma)$ shifts to
higher densities and $\Delta v(\gamma)$ decreases, eventually vanishing
for some randomness $\gamma_c$. The point of density $n_c=n_t(\gamma_c)$
and
average velocity $v_c=v(n_c)$, belonging to the curve for $\gamma_c$,
will correspond to a critical point. This conclusion is further
supported
by the large increase of the fluctuations of $v$ observed in the
vicinity of that
point. As can be inferred from Fig.\ \ref{Isotermas}, the location of
the critical point lies somewhere in the range
$0.45\leq\gamma_c\leq 0.5$, but it cannot be more accurately determined
from our simulations first, because $\gamma$ is an input parameter, and
second, due to the strong size-dependence of $\gamma_c$.

The part of the curves $v(n)$ corresponding to the free phase (which for
$\gamma=1/2$ means the whole curve) fits rather well the linear law
$v(n)=(1-n)/2$. It can be proven analytically for the infinite system
\cite{7} that this is precisely the asymptotic behavior of $v(n)$ when
$n\to 0$, for any value of $\gamma$. It is
remarkable that the agreement with the simulations is rather good even
far from this limit.

The jammed phase, and in fact the nature of the transition, can be
better understood by
analyzing the average distribution of cars on the lattice. Fig.\
\ref{Isotermas} shows in this phase, for the lowest values of
$\gamma$, a few small jumps in which the value of $v$
increases. The explanation of these jumps is the following: Before the
jamming transition occurs, cars are distributed homogeneously, whereas
after the transition cars always order along broad diagonal strips
extending throughout the whole system (see Fig. \ref{multistrip}),
with the two types of cars roughly separated in two halves.
These strips do not trap empty sites (holes) inside; thus,
the observed remnant average velocity is due only to the movement of
the cars on the borders of the strips. Different number of strips
characterize different ordered phases. A given initial configuration
goes to one of this phases with a certain probability. For a
given density, we compute this
probability by taking a large number of initial configurations
and counting how many of them go to each phase.
The stable phase will be
that of maximum probability, the rest of them being metastable.
Accordingly, in Fig.\ \ref{Isotermas} we plot the velocity of the stable
phase. The small jumps correspond to a exchange of stability between two
phases. In these jumps $v$ increases since every new strip provides
two more borderlines along which cars can move. As this remnant
movement is just a ``surface" effect,
it should vanish when $L\to\infty$; however, at the same time the number
of strips increases, hence supplying extra moving cars. The resulting
value in the infinite system will depend on this competition of effects;
we will return to this point later on.

For $\gamma=0$ the results are similar to those reported in
\cite{5} (the only difference being that our velocities are, by
definition, half theirs). Since cars do not change direction, any
initial configuration ends up  either in a periodic or
in a stuck ($v=0$) state.
The predominance of each kind of state decides in which
phase is the system.
According to the authors of \cite{5}, their results
do not allow them to exclude the possibility that
$n_t(0)\to 0$ as $L\to\infty$. In contrast,
though in our simulations the values of $n_t(\gamma)$ also decrease
as $L$ increases,
we can clearly see that for the lowest values of $\gamma$ (say,
$\gamma=0.1$, 0.2 and 0.3) it already
converges to a nonzero value, even for the relatively small sizes we
are dealing with. Besides, as we have commented on above,
we have proven \cite{7} that the slope of $v(n)$, for infinite
$L$, is exactly $-1/2$, independently of $\gamma$,
in the limit $n\to 0$.
This result also holds for $\gamma=0$;
however, the simulations of \cite{5} indicate that the value of the
slope in the free phase is $0$ for $L$ up to $512$.
This fact supports the idea that
when $L\to\infty$, $n_t(0)\to 0$, though the possibility of a change
of slope from $0$ to $-1/2$ at much larger system sizes is not excluded,
but seems very unlikely.

{}From the size-dependence of the parameters of our simulation we can
draw an image of what happens in
model A when $L\to\infty$. On the one hand,
as we have already pointed out, $n_t(\gamma)$ converges to a
nonzero value in the infinite system.
On the other hand, the $\gamma=0.5$ curve
does not change with $L$, and the critical point moves towards this
curve as $L$ increases (it cannot be inferred from our results whether
$\gamma_c$ finally reaches the value 1/2). Accordingly, even though the
transition densities, $n_t(\gamma)$, for the rest of the values of
$\gamma$ still decrease, they should
reach a nonzero value when $L\to\infty$. This part of the phase
diagram will thus not qualitatively change at infinite size.
Regarding the structure of the jammed phase, as
strips never trap holes, the only possible
way that such structures survive in an infinite city with $n\ne 0$
and $n\ne 1$ is that an infinite number of strips appear.
Consequently, those ``transitions" between jammed phases with different
 number
of strips are a finite system size effect; a result that is further
supported by the fact that such transitions move quickly towards
$n_t(\gamma)$ as $L$ increases.
The infinite system will then be formed by infinite strips with
a typical size and a typical separation (which will in general depend
on $n$), and the value of $v$ will
simply be the ratio of the average number of moving cars per strip to
the average number of cars per strip. Nevertheless, as the
average separation between strips seems comparable to the sizes used in
our simulations, the values of $v$ obtained are still affected by strong
finite-size effects.

The phase diagram of model B is similar to that of model A. There
is also a phase transition from the ``free" to the jammed
regime, with diagonal strip structure right after the transition.
This is indicated as before, by a sharp decrease of the velocity
for a certain value of the density $n_t(\gamma )$ (smaller than in
model A).
The dependence of the parameters characterizing the transition,
$n_t(\gamma )$ and $\Delta v(\gamma)$, on $\gamma$ and the size of the
city
$L$ is qualitatively the same as in model A.
The main differences of this model
are in the structure of the jammed states appearing after the
transition. Having this model four different types of cars
and streets, it has a symmetry (absent in model A) under $90^\circ$
rotations, and the jammed strip can appear with equal
probability along each diagonal direction.
Besides that, the strip has inside some holes that
allow the diffusion through the jam of the different car types.
They also contribute to the remnant velocity in the jam state
and could have
some significant effect in the infinite size case.
The other important departure from the behavior of model A is the
type of stable phases present in the jammed region.
First of all, while in model A the jammed phase can show
multiple strips, in model B we only see one strip.
This strip is composed of two longitudinal halves, each
containing a mixture of two types of cars.
For example, if the strip runs from the lower-left to the upper-right,
the upper half of it is mainly composed
of cars of the types trying to go right and down,
and the bottom half by the ones trying to go left and up.
Secondly, as density is increased, a point is reached where the
form  of the stable jammed phase suddenly changes. The
majority of the holes, that before this point were forming a paralell
strip to the cars, now arrange themselves in a closed square-like
regions (see Fig.\ \ref{emptyholes}). Meanwhile, the cars
in the jam have separated in four regions according to their type.
The cars trying to go up are above each empty region, to the left
the ones trying to go left, and so on.
This change of structure produces a noticeable, though small,
change in the slope
of the velocity curves $v(n)$ in the phase diagram.
Our simulations do not allow us to conclude whether this transition is
continuous or weakly first order.
More work on this point is in progress \cite{7}.

{}From our data we can only conjecture what the structure of the jammed
phase in model B  will be, as the size of system goes to infinity.
We have already seen, in some preliminary runs, that the
number of those empty regions increases, as we simulate, at constant
density, in larger cities. Keeping the parallelism with
model A, we can think that in the limit of infinity size an infinite
number of strips will appear (though the existence of holes inside
them weakens this conclusion). However, due to the
90$^{\circ}$-rotation symmetry
present in model B, the strips may appear in both diagonal directions
simultaneously. If this happened the stable state would be one in which
there would be an infinite number of square-like empty regions, as
described above, arranged in a kind of
lattice structure. We hope that we will be able to settle this question
in the future.

In summary, we have studied models incorporating what we think are
the essential ingredients (excluded volume and turn capability)
of urban car movement in cities with realistic
structures (defined by the arrangement of the streets and the
organization of traffic lights) and we have found a first order phase
transition from a freely moving regime to a jammed state.
It is important to check whether this striking feature still stands
when more elements are added (say, disorder via
forbidden streets or non-synchronized
traffic-lights, preferred streets, rush hours, etc).
If this happens, it will be
possible to conclude that, as a consequence of car interaction,
any city will {\em always} have a saturation
density of cars after which the average velocity falls sharply.
This could be an important issue to consider in the design of city
and traffic policies. If a city, with a given density of cars, were
saturated, local improvements would have little effect
on the average velocity, and only global changes in
the city capacity or the number of
cars will be able to solve the
problem. More analytical and numerical work is currently being developed
\cite{7} along this line.

Two of us (J.A.C. and A.S.) acknowledge financial support of two
projects (PB91-0378 and MAT90-0544, respectively) of the Direcci\'{o}n
General de Investigaci\'{o}n Cient\'{\i}fica y T\'{e}cnica (Spain).

\begin{figure}
\caption[]{Average velocity, $v$, for cars in model A as a function of
car density, $n$, for different randomness values, $\gamma$, in a city
of $64\times 64$ streets. The full lines are a guide to the eye. The
dashed line is an approximate fit to the points where the transition
occurs.}
\label{Isotermas}
\end{figure}
\begin{figure}
\caption[]{Average site density of stopped cars in model A
($128\times 128$ streets) in the jammed phase, for a car density $n=0.7$
and a randomness $\gamma=0.1$. Different values are represented by
different grey levels ranging from black (site always empty) to white
(site always occupied). This picture illustrate the multistrip structure
of the jammed phase in model A.}
\label{multistrip}
\end{figure}
\begin{figure}
\caption[]{Same as Fig.\ \ref{multistrip}, for model B and car density
$n=0.9$ and randomness $\gamma=0.2$. This parameters correspond to the
second ordered phase (see text).
The existence of closed empty regions
surrounded by cars is illustrated.}
\label{emptyholes}
\end{figure}

\end{document}